\documentclass[preprint,aps,floats]{revtex4}
\usepackage{epsfig}
\usepackage{graphicx}
\usepackage{dcolumn}
\usepackage{bm}
\def\lsim{\mathrel {\vcenter {\baselineskip 0pt \kern 0pt
    \hbox{$&lt;$} \kern 0pt \hbox{$\sim$} }}}
\def\gsim{\mathrel {\vcenter {\baselineskip 0pt \kern 0pt
    \hbox{$&gt;$} \kern 0pt \hbox{$\sim$} }}}
\newcommand{\U}{{\cal {U}}}

\begin{document}

\title{Constraints on Unparticle Interactions from \\Particle and Antiparticle  Oscillations}
\author{Shao-Long Chen$^1$, Xiao-Gang He$^{1,2}$, Xue-Qian Li$^2$, Ho-Chin Tsai$^1$ and Zheng-Tao Wei$^2$}
\affiliation{ $^1$Department of Physics and Center for Theoretical
Sciences, National Taiwan University, Taipei\\
$^2$Department of Physics, Nankai University, Tianjin 300071}

\date{\today}

\begin{abstract}
We study unparticle effects on particle and antiparticle
osillations in meson-antimeson, and muonium-antimuonium systems.
Unlike usual tree level contributions to meson oscillations from
heavy particle exchange with small $\Gamma_{12}$, the unparticle
may have sizeable contributions to both $M_{12}$ and $\Gamma_{12}$
due to fractional dimension $d_\U$ of the unparticle. We find that
very stringent constraints on the unparticle and particle
interactions can be obtained. If unparticle effect dominates the
contributions (which may happen in $D^0-\bar D^0$ mixing) to meson
mixing parameters $x$ and $y$, we find that $x/y =\cot(\pi d_\U)$.
Interesting constraints on unparticle and particle interactions
can also be obtained using muonion and antimuonion oscillation
data. We also comment on unparticle effects on CP violation in
meson oscillations.
\end{abstract}


\maketitle

\section{Introduction}

Recently Georgi proposed an interesting idea to describe possible
scale invariant effect at low energies by
unparticles\cite{Georgi:2007ek}. Georgi argued that operators
$O_{BZ}$ made of BZ fields in the scale invariant sector may
interact with operators $O_{SM}$ of dimension $d_{SM}$ made of
Standard Model (SM) fields at some high energy scale by exchange
particles with large masses, $M_{\cal{U}}$, with the generic form
$O_{SM} O_{BZ}/M^k_{\cal{U}}$. At another scale
$\Lambda_{\cal{U}}$ the BZ sector induce dimensional
transmutation, below that scale the BZ operator $O_{BZ}$ matches
onto unparticle operator $O_{\cal{U}}$ with dimension $d_\U$ and
the unparticle interaction with SM particles at low energy has the
form \begin{eqnarray} \lambda \Lambda_{\cal{U}} ^{4-d_{SM} - d_\U}
O_{SM} O_{\cal{U}}.
\end{eqnarray}

Study of unparticle effects has drawn a lot of attentions from
collider physics~\cite{Georgi:2007ek,collider}, low energy flavor
conserving and flavor violating
processes~\cite{lowenergy,b-mixing,flavor1}, long range
effects~\cite{longrange}, cosmological and astrophysics
phenomena~\cite{astro}, to more theoretical studies~\cite{theory}.
In this work we further study unparticle effects on particle and
antiparticle oscillations of meson $P$ and antimeson $\bar P$, and
muonium and antimuonium systems.

Unparticle effects on oscillation in meson and antimeson have been
considered previously~\cite{b-mixing}. Our investigation for meson
and antimeson oscillation will focus on some interesting features
due to the fractional dimension of unparticle $d_\U$. Unlike usual
tree level contributions to meson oscillations from heavy particle
exchange with small $\Gamma_{12}$, the unparticle may have
sizeable contributions to both $M_{12}$ and $\Gamma_{12}$ due to
fractional dimension $d_\U$ of the unparticle leading to a phase
factor $(-1)^{d_\U -2}$ in the propagation.  If unparticle effect
dominates the contributions (which may happen in $D^0-\bar D^0$
mixing) to meson mixing parameters $x$ and $y$, we find that $x/y
=\cot(\pi d_\U)$.

Meson-antimeson oscillation exists in several neutral meson
systems, $K^0 - \bar K^0$, $D^0 - \bar D^0$, $B_d^0 - \bar B^0_d$
and $B_s^0 - \bar B_s^0$. Long distance contributions to
oscillation parameters for $K^0 -\bar K^0$ are large which causes
a large uncertainty in theoretical calculations. We will restrict
our calculations  for unparticle effects to $D^0-\bar D^0$ and
$B_{d,s}^0 - \bar B^0_{d,s}$ systems. We find that very stringent
constraints on the unparticle and particle interactions can be
obtained.

Muonium ($M=(\bar \mu e$)) and antimuonium ($\bar M = (\bar e \mu
)$) oscillation may also provide interesting constraints on flavor
changing interaction. Experimentally muonium-antimuonium
oscillation has not been established. Our analysis shows that
constraints on unparticle and particle interactions can indeed be
obtained using experimental data on muonium-antimuonium
oscillation.

\section{Meson and Antimeson Oscillations}

The mixing of a meson and its antimeson is determined by the off
diagonal matrix elements $M_{12}$ and $\Gamma_{12}$ in the
Hamiltonian. Their relations to the mass and lifetime differences
are given by,
\begin{eqnarray}
&&(m_H - m_L)-i (\Gamma_H-\Gamma_L)/2 =2
\sqrt{(M_{12}-i\Gamma_{12}/2)(M_{12}^* - i\Gamma^*_{12}/2)},
\end{eqnarray}
where the subscripts ``H'' and ``L'' label the mass eigenstates,
$\vert P_H \rangle = p\vert P\rangle + q \vert \bar P\rangle$
and $\vert P_L \rangle = p\vert P\rangle - q \vert \bar P\rangle$,
respectively. $p$ and $q$ are normalized as $\vert p\vert^2 +
\vert q \vert^2 =1$ and $({q/ p})^2 = (M_{12}^*-i\Gamma^*_{12}/2)/
(M_{12}-i\Gamma_{12}/2)$. We denote the mass and lifetime
differences by $\Delta m = m_H - m_L$ and $\Delta \Gamma =
\Gamma_H - \Gamma_L$. The parameters $x$ and $y$ are related to
$\Delta m$ and $\Delta \Gamma$ by $x=\Delta m/\Gamma$ and $y =
\Delta \Gamma/2\Gamma$.

There are several possible contributions to $M_{12}$ and
$\Gamma_{12}$ from unparticle and particle interactions. The
following operators composed of SM fields and derivatives with
dimensions less than or equal to 4 invariant under the SM gauge
can contribute to meson mixing at tree level,
\begin{eqnarray}
a): &&\lambda'_{QQ}\Lambda_{\cal{U}}^{1-d_\U}\bar Q_L \gamma_\mu
Q_L O^\mu_{\cal{U}}, \;\lambda'_{UU}\Lambda_{\cal{U}}^{1-d_\U}\bar
U_R \gamma_\mu U_R O_{\cal{U}}^\mu,
\;\lambda'_{DD}\Lambda_{\cal{U}}^{1-d_\U}\bar D_R  \gamma_\mu D_R O^\mu_{\cal{U}};\nonumber\\
b): &&i\lambda_{QQ}\Lambda_{\cal{U}}^{-d_\U} \bar Q_L \gamma_\mu
D^\mu Q_L O_{\cal{U}}, \;i\lambda_{UU}\Lambda_\U^{-d_\U}\bar
U_R\gamma_\mu D^\mu U_R O_{\cal{U}},
\;i\lambda_{DD}\Lambda_{\cal{U}}^{-d_\U}\bar D_R \gamma_\mu D^\mu D_R O_{\cal{U}};\nonumber\\
c): &&i\tilde \lambda_{QQ}\Lambda_{\cal{U}}^{-d_\U}\bar Q_L
\gamma_\mu Q_L
\partial^\mu O_{\cal{U}}, \;i\tilde \lambda_{UU}\Lambda_{\cal{U}}^{-d_\U}\bar U_R \gamma_\mu
U_R \partial^\mu O_{\cal{U}},
\;i\tilde \lambda_{DD}\Lambda_{\cal{U}}^{-d_\U}\bar D_R  \gamma_\mu D_R
\partial^\mu O_{\cal{U}};\nonumber\\
d): &&\lambda_{YU}\Lambda_{\cal{U}}^{-d_\U}\bar Q_L H U_R
O_{\cal{U}}, \;\lambda_{YD}\Lambda_{\cal{U}}^{-d_\U}\bar Q_L
\tilde H D_R O_{\cal{U}}.
\label{operator}
\end{eqnarray}
Here $Q_L$, $U_R$, and $D_R$ are the SM left-handed quark doublet,
right-handed up-quark, and right-handed down-quark, respectively.

\begin{figure}[t!]
\includegraphics[width=5.5 in]{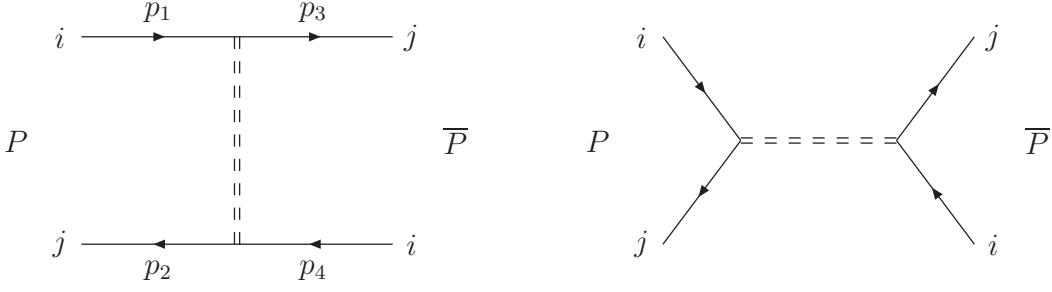}
\caption{\label{feynman} \small The $t$ and $s$ channel
contributions to meson-antimeson oscillation.}
\end{figure}

After using equation of motion for quarks, the interactions in
eq.(\ref{operator}) can be parameterized in the following form
\begin{eqnarray}
&&\mbox{For vector}\;\;O^\mu_\U: {\cal L}_V = \Lambda_\U^{1-d_\U}\bar q_i
(c_{V_L}^{ij}\gamma_\mu L + c_{V_R}^{ij}\gamma_\mu R) q_j O^\mu_\U\;;
\nonumber\\
&&\mbox{For scalar}\;\;O_\U: {\cal L}_S =
\Lambda_\U^{1-d_\U}{m_j\over \Lambda_\U} \bar q_i (c_{S_L}^{ij} L
+ c_{S_R}^{ij} R) q_j O^\mu_\U\;,
\end{eqnarray}
where $q_{i(j)}$ denote quarks with flavor indices $i(j)$. The
parameters $c_{V_{L,R}}^{ij}$ and $c_{S_{L,R}}^{ij}$ are different for the
interactions listed in a) to d). They are given by
\begin{eqnarray}
a): &&c^{ij}_{V_L} = \lambda_{QQ}^{'ij},\;\;c^{ij}_{V_R} =
\lambda^{'ij}_{UU,DD};\nonumber\\
b):&&c^{ij}_{S_L} = {1\over m_j}(\lambda^{ji*}_{QQ}m_i
+\lambda^{ij}_{UU,DD} m_j)
,\;\;c^{ij}_{S_R} = {1\over m_j}(\lambda^{ij}_{QQ} m_j
+\lambda^{ji*}_{UU,DD}m_i);\nonumber\\
c):&&c^{ij}_{S_L} = {1\over m_j}(\tilde \lambda^{ij}_{QQ}
m_i - \tilde \lambda^{ij}_{UU,DD}m_j),\;\;c^{ij}_{S_R} = {1\over m_j}( \tilde
\lambda^{ij}_{UU,DD}m_i-\tilde
\lambda^{ij}_{QQ} m_j );\nonumber\\
d):&&c^{ij}_{S_L} = v \lambda^{ji*}_{YF},\;\;c^{ij}_{S_R} = v
\lambda^{ij}_{YF}.
\end{eqnarray}
where $v = \langle H\rangle$ is the vacuum expectation value of
$H$. Note that the vector unparticle couplings $c_V$ scaled as
$\Lambda_\U^{1-d_\U}$, while scalar unparticle couplings $c_S$
scaled as $\Lambda_\U^{-d_\U}$.

Evaluating the two diagrams in Fig.~\ref{feynman}, we obtain
\begin{eqnarray}
\mbox{For}&& O^\mu_\U: \nonumber\\
&&H^\U_{eff} = - {A_{d_\U}\over 2 \sin(\pi d_\U)}
\Lambda_\U^{2(1-d_\U)} e^{-i\pi d_\U}\left \{{1\over
4} \left ( {1\over
(s)^{2-d_\U}} + {1\over (t)^{2-d_\U}}\right )\left (\bar q_i
(c^{ij}_{V_L}\gamma_\mu L + c^{ij}_{V_R}\gamma_\mu R) q_j\right
)^2\right .\nonumber\\&&\;\;\;\;\;\;\;\; \left .+ {1\over
4} \left ( {1\over
s(s)^{2-d_\U}} + {1\over t(t)^{2-d_\U}}\right )\left (\bar q_i
(c^{ij}_{V_L}(m_iL -m_jR)+ c^{ij}_{V_R}(m_iR -m_jL)) q_j\right
)^2\right\}
;\nonumber\\
\mbox{For} &&O_\U: \nonumber\\
&&H^\U_{eff} = {A_{d_\U}\over 2 \sin(\pi
d_\U)}\Lambda_\U^{2(1-d_\U)}{m^2_j\over \Lambda_\U^2} e^{-i\pi
d_\U} {1\over
4} \left ( {1\over (s)^{2-d_\U}} + {1\over (t)^{2-d_\U}}\right
) \left (\bar q_i (c^{ij}_{S_L}L + c^{ij}_{S_R}R) q_j\right
)^2.\label{matrix}
\end{eqnarray}
Here $A_{d_\U} =
(16\pi^{5/2}/(2\pi)^{2d_\U})\Gamma(d_\U+1/2)/(\Gamma(d_\U
-1)\Gamma(2d_\U))$. We have used $(iA_{d_\U}/2\sin(\pi
d_\U))\times (1/(-p^2)^{2-d_\U})$ and $(iA_{d_\U}/2\sin(\pi
d_\U))\times ((-g^{\mu\nu} + p^\mu p^\nu/p^2)/(-p^2)^{2-d_\U})$
for scalar and vector unparticle propagators, respectively.

In the systems we are studying, mesons are made of a light
(labelled by i) and a heavy quark (labelled by j). In the heavy
quark limit, one has $s = t\approx m^2_j \approx m^2_P$. With this
approximation and theoretical matrix elements for the relevant
operators, we have
\begin{eqnarray}
&&M^\U_{12}= A_{d_\U} \left ({m^2_P\over \Lambda_\U^2}\right
)^{d_\U-1} {f^2_P\over 12 m_P} \left \{ -(c^{ij}_V)^2 (B_V
-{5\over 8}B_S)  -
{m^2_P\over \Lambda_\U^2}(c^{ij}_S)^2 {5\over 8} B_S \right \}\cot(d_\U\pi),\nonumber\\
&&\nonumber\\
 &&\Gamma_{12}^\U = 2 M^\U_{12}\tan(d_\U\pi).\label{ff}
\end{eqnarray}
We have included a missing factor of 1/2! due to Wick rotation in
previous studies~\cite{b-mixing}. The parameters $B_{V,S}$ are the
bag factors which are equal to 1 in the vacuum saturation and
factorization approximation.

We would like to point out some silent features of the unparticle
contribution to $M_{12}^\U$ and $\Gamma_{12}^\U$ due to the phase factor
$e^{-i\pi d_\U}$. We note that $M_{12}^\U$ can have both sign
depending on the value of $d_\U$ due to the factor $\cot(\pi
d_\U)$, therefore if information about the sign can be obtained
from other considerations, the dimension $d_\U$ can be restricted.
There may be a sizeable contribution to $\Gamma_{12}$ at tree
level which is not possible for usual tree level heavy  particle
exchange. For $d_\U$ equal to half integers, there is no
contribution to $M_{12}$, but there is for $\Gamma_{12}$. Another
interesting feature of unparticle contribution is that the ratio
$M_{12}/(\Gamma_{12}/2)$ of unparticle contribution is related to
the unparticle dimension parameter $d_\U$ by
\begin{eqnarray}
{M_{12}^\U\over \Gamma_{12}^\U/2} = \cot(\pi d_\U).
\end{eqnarray}
If the unparticle contribution dominates meson and antimeson
oscillation then the measurements of $M_{12}$ and $\Gamma_{12}$
provide a possible way to determine the dimension parameter
$d_\U$.

\begin{figure}[thb!]
\includegraphics[width=5.5 in]{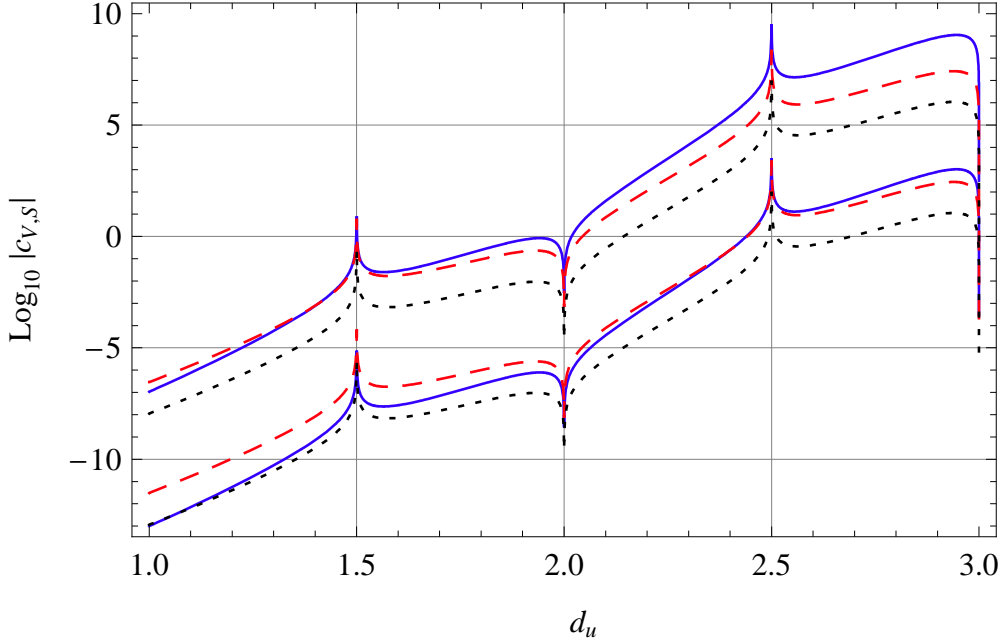}
\caption{\label{dm} \small Constraints on the coupling $c_S$
(upper curves) and $c_V$ (lower curves) as functions of $d_\U$
from meson-antimeson oscillation with $\Lambda_\U = 1$ TeV by
fitting the mass differences $\Delta m$. The solid (blue), dashed
(red), and dotted (black) curves are for $D^0-\bar D^0$, $ B_d-
\bar B_d$ and $ B_s-\bar B_s$, respectively.}
\end{figure}

\begin{figure}[thb!]
\includegraphics[width=5.5 in]{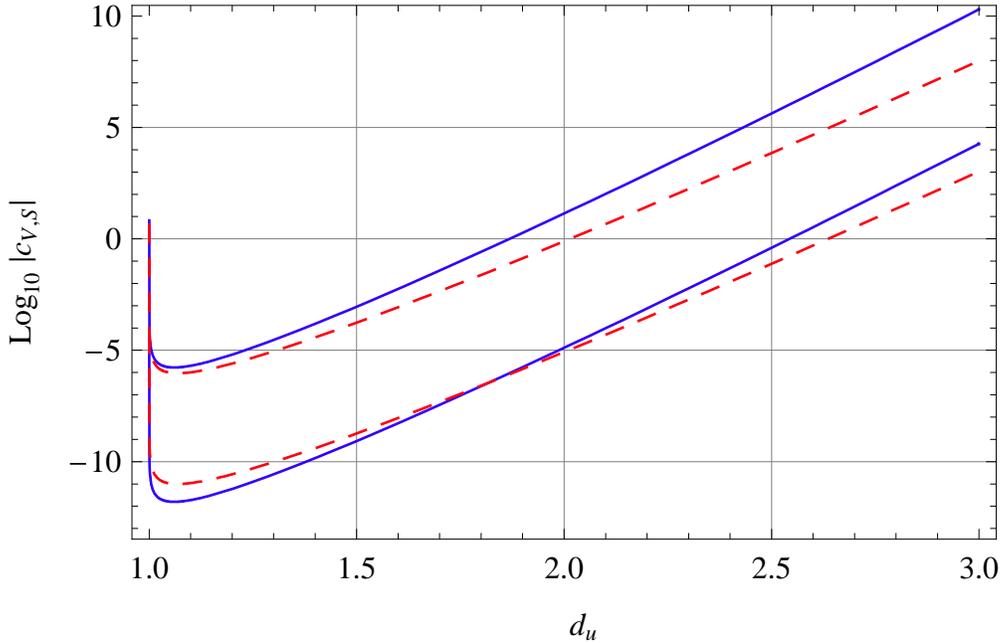}
\caption{\label{dg} \small Constraints on the coupling $c_S$
(upper curves) and $c_V$ (lower curves) from meson-antimeson
oscillation with $\Lambda_\U = 1$ TeV by fitting the width
differences $\Delta \Gamma$. The solid (blue) and dashed (red)
curves are for $D^0-\bar D^0$ and $B_s-\bar B_s$, respectively.}
\end{figure}

We now present our numerical results on the constraints for
unparticle and particle interactions with the assumption of CP
conservation. In this case the unparticle contribution to mass
difference $\Delta m^\U = 2|M_{12}^\U|$ and $x^\U/y^\U =
M^\U_{12}/(\Gamma^\U_{12}/2) = \cot(\pi d_\U)$. We will comment on
possible CP violating effects later. The results are shown in
Figs.~\ref{dm} and~\ref{dg}. The constraints are obtained with the
vacuum saturation approximation, i.e. $B_V = B_S = 1$ and $f_D =
0.201$~GeV, $f_{B_d} = 0.216$~GeV and $f_{B_s} =
0.260$~GeV~\cite{Okamoto:2005zg}.

\noindent{\bf $D^0 - \bar D^0$ System}

Belle and BABAR collaborations have recently published evidence
for $D^0-\bar D^0$ oscillation~\cite{Staric:2007dt}.  The Heavy
Flavor Averaging Group (HFAG)~\cite{hfag_charm} combined all
mixing measurements to obtain world average (WA) values for $x$ and
$y$ for CP conserving case with $x=(0.87^{+0.30}_{-0.34})\%$ and
$y=(0.66^{+0.21}_{-0.20})\%$. Short distance contributions in the SM
are several orders of magnitudes smaller than the central
experimental values. There are possible large long distance
contributions which are however difficult to have precise
predictions. We will assume that the contributions to $x$ and $y$
are purely from unparticle effects. With this assumption, we
immediately obtain
\begin{eqnarray}
\frac{x}{y} = \cot(\pi d_\U)\simeq 1.31 \pm 0.61\;.
\end{eqnarray}

Although the sign of $x$ and $y$ are both positive, the absolute
sign of $M_{12}^\U$ cannot be determined from $x$ measurement,
therefore $\pi d_\U$ can be in the first and third quadrants from
the sign of $x/y$. We have
\begin{equation}
d_\U=(0.21+n) \pm 0.07\;,
\end{equation}
where $n$ is an integer number which cannot be determined from
just information from $x/y$. Note that the experimental errors are
still large, consequently the uncertainties of $d_\U$ are
substantial.

One can also obtain constraints on the couplings $c_{V_{L,R}}$ and
$c_{S_{L,R}}$ from $x$ or $y$ for different $d_\U$, allowing
unparticle contributions to saturate the experimental upper bound
on $x_D$ or $y_{D}$. As the contributions from unparticle are
suppressed by factors of $(m^2_D/\Lambda^2_\U)^{d_\U -1}$ and
$(m^2_D/\Lambda^2_\U)^{d_\U}$ for vector and scalar unparticle
respectively,  if  $n$ is large the contributions are negligible
for a fixed $\Lambda_\U$. We will therefore just consider the
lowest possibilities with phase $\pi d_\U$ covering all four
quadrants plotting results for $d_\U$ in the range of 1 to 3 for
illustration in Fig.~\ref{dm} (solid curves) with $\Lambda_\U$
fixed to be 1 TeV. In this range, the phase $\pi d_\U$ will cover
all four quadrants. At $d_\U$ equal to half integers (1.5, 2.5),
there is no contribution to $\Delta m^\U$, and therefore there
are no constraints on $c_{V,S}$. This is indicated by the two
peaks at $d_\U =1.5$ and $2.5$ in Fig.~\ref{dm}. At $d_\U$ equal to
1, $\sin (\pi d_\U) =0$, naively the contribution blows off.
However, at $d_\U = 1$, $A_{d_\U}/\sin(\pi d_\U)$ is finite and
therefore there is finite contribution as should be for a
dimension one particle. For other integers, the contribution blows
off. For these reasons, when reading Fig.~\ref{dm}, one should not
taken values too close to integers larger than 1 and half integers
for $d_\U$. One can also use data on $y_D$ to constrain $c_{V,S}$.
The results are shown in Fig.~\ref{dg} (solid curve). Note that in
this case, at $d_\U =1$, there is no contribution to $y_D$ because
$A_{d_\U} =0$.

From Figs.~\ref{dm} and \ref{dg} (solid curves), we see that
constraints from $x_D$ and $y_D$ give similar constraints for
$c_{V,S}$ when away from integers and half-integers. The
constraint for $c_S$ is weaker than $c_V$ because of the relative
suppression factor $m^2_D/\Lambda_\U^2$ as pointed out earlier.
Smaller $d_\U$ give stronger constraints on $c_{V,S}$.

\noindent{\bf $B^0_{d} - \bar B^0_{d}$ System}

$\Delta m_{B_d}$ has been measured to be $(0.507 \pm
0.005)\mbox{ps}^{-1}$ ($x = 0.776\pm 0.008$) \cite{Yao:2006px,
HFAG}. A non-zero $\Delta\Gamma_{B_d} (y)$ has not been
established. The SM prediction for $\Delta m_{B_d}$ agrees with
data very well. The prediction for $\Delta \Gamma_{B_d}$
is $-(26.7^{+5.8}_{-6.5})\times 10^{-4}\mbox{ps}^{-1}$
~\cite{LN2006} which is very small to be measured experimentally. Since SM
prediction for $\Delta m_{B_d}$ agrees with data, in our analysis
to obtain constraints on the couplings $c_{V,S}$ we will just
allow the unparticle contributions to vary within 2$\sigma$ of
experimental error bar.

Since the SM prediction for $M_{12}^{SM}$ is positive, if the
unparticle contribution is required to increase (decrease) the
value for $\Delta m_{B_d}$ relative to the SM value, then $\pi
d_\U$ needs to be in second and fourth (first and third)
quadrants. In Fig.~\ref{dm}, we show the results for constraints
on $c_{V,S}$ assuming the unparticle contributions saturate
$2\sigma$ of experimental error bar for $\Delta m_{B_d}$. The
constraints are similar to those obtained from $D^0 - \bar D^0$
oscillation. The predicted value for $y^\U_{B_d}$ is given by
$y^\U_{B_d} = x^\U_{B_d} \tan(\pi d_\U)$ which can be as large as
present experimental upper bound since at $d_\U$ close to half
integers $\cot(\pi d_\U)$ can be very large, and at half integers
there is no constraint from $\Delta m$. Future experiments may
tell us more.

\noindent{\bf $B^0_{s} -\bar B_s^0$ System}

For $B_s^0 - \bar B_s^0$, $\Delta m_{B_s}$ is measured to be
$(17.77\pm 0.12)\mbox{ps}^{-1}$~\cite{deltamsexp}, and  $\Delta
\Gamma_{B_s} = -(0.084^{+0.055}_{-0.050})\mbox{ps}^{-1}$
\cite{Yao:2006px,HFAG}. SM best fits are $\Delta m_{B_s} =
(19.3\pm6.68)\mbox{ps}^{-1}$ and $\Delta \Gamma_{B_s} = -(0.096\pm
0.039)\mbox{ps}^{-1}$ ~\cite{LN2006}. There are differences for
central values of SM predictions and experimental measurements.
Taking these central values and attributing the differences are due
to unparticle effects, one would favor $\pi d_\U$ to be in the
first or third quadrants. Since both SM predictions and
experimental measurements have large errors and they agree with in
error bars, we will present our constraints on $c_{V,S}$ taking,
again, $2\sigma$ experimental error bars for both $\Delta m$ and
$\Delta \Gamma$. The results are shown in Figs.~\ref{dm}
and~\ref{dg}. The constraints on $c_{V,S}$ are similar to those
obtained from $D^0 - \bar D^0$ oscillation.

\section{Muonium-antimuonium Oscillation}

For muonium and antimuonium oscillation to occur, there must be
flavor changing interactions. To the lowest order, the following
unparticle and particle interaction operators will contribute,
\begin{eqnarray}
a): &&\lambda'_{LL}\Lambda_{\cal{U}}^{1-d_\U}\bar L_L  \gamma_\mu
L_L O^\mu_{\cal{U}}, \;\lambda'_{EE}\Lambda_{\cal{U}}^{1-d_\U}\bar
E_R\gamma_\mu E_R O^\mu_{\cal{U}};\nonumber\\
b):&&i\lambda_{LL}\Lambda_{\cal{U}}^{-d_\U}\bar L_L \gamma_\mu
D^\mu L_L O_{\cal{U}}, \;i\lambda_{EE}\Lambda_{\cal{U}}^{-d_\U}
\bar E_R\gamma_\mu D^\mu E_R O_{\cal{U}};\nonumber\\
c): &&i\tilde \lambda_{LL}\Lambda_{\cal{U}}^{-d_\U}\bar L_L
\gamma_\mu L_L \partial^\mu O_{\cal{U}}, \;i\tilde
\lambda_{EE}\Lambda_\U^{-d_\U}\bar E_R \gamma_\mu E_R
\partial^\mu O_{\cal{U}};\nonumber\\
d): &&\lambda_{YE}\Lambda_{\cal{U}}^{-d_\U}\bar L_L\tilde H E_R
O_{\cal{U}}. \label{operator-lep}
\end{eqnarray}

At tree level, exchange of unparticles will generate $\bar \mu
\Gamma_1 e \bar \mu \Gamma_2 e$ type of matrix elements. The
operators has the same form given in eq.(\ref{matrix}) with
appropriate replacements of quarks by letptons and the associated
couplings.

The SM prediction for muonium and antimuonium oscillation is
extremely small. Observation of this oscillation at a
substantially larger rate will be an indication of new physics.
Experimentally, no oscillation has been observed. The current
upper limit for the probability of spontaneous muonium to
antimuonium conversion was established at $P_{\bar MM}\leq
8.3\times 10^{-11}$ (90\% C.L.) in 0.1 T magnetic field
\cite{Willmann:1998gd}.

In the absence of external electromagnetic fields, the probability
$P_{\bar MM}$ of observing a transition  can be written
as~\cite{formula} $P_{\bar MM}(0 \mbox{T})\simeq
{|\delta|^2}/{(2\Gamma_\mu^2)}$, where $\delta\equiv 2\langle\bar
M|H_{eff}|M\rangle$ and $\Gamma_\mu$ is the muon decay width. Here
the effective Hamiltonian is defined as $H_{eff} = (G_{\bar
MM}/\sqrt{2}) \bar \mu\Gamma_1 e\bar \mu\Gamma_2 e$. For the
$\Gamma_1 \times \Gamma_2 =(V\pm A)^2$ type Hamiltonian
$(\bar\mu\gamma_\lambda(1-\gamma_5)e)^2$, the transition amplitude
is given by $\delta=16G_{\bar MM}/(\sqrt{2}\pi a^3)$  for both
triplet and singlet muonium states, where $a\simeq (\alpha
m_e)^{-1}$ is the Bohr radius. But for $(S\pm P)^2$ type we have
$\delta=-4G_{\bar MM}/(\sqrt{2}\pi a^3)$ for both triplet and
singlet muonium~\cite{Hou:1995dg}.

As for our case, omitting $m_e$, the contributions corresponding
to parameters $c_{V_{L,R}}$ are $(V\pm A)^2+(S\pm P)^2$ type and
to parameters $c_{S_{L,R}}$ are $(S\pm P)^2$ type. Therefore we
have $\delta$ given by $\delta=12G_{\bar MM}/(\sqrt{2}\pi a^3)$ and
$\delta=-4G_{\bar MM}/(\sqrt{2}\pi a^3)$ for
$c_{V_{L,R}}$ and $c_{S_{L,R}}$ respectively.

It is important to note that the probability $P_{\bar MM}$ has
strong magnetic field dependence which usually occurs in
experimental situation. With an external magnetic field, there is
a reduction factor $S_B$, i.e. $P_{\bar MM}(B)=S_B P_{\bar MM}(0
\mbox{T})$. The magnetic field correction factor $S_B$ describes
the suppression of the conversion in the external magnetic field
due to the removal of degeneracy between corresponding levels in
$\bar M$ and $M$. One has $S_B=0.35$ for $(V\pm A)^2$ and $(S\pm
P)^2$ type interactions at $B=0.1 \mbox{T}$~\cite{Willmann:1998gd,
Horikawa:1995ae}. Using this experimental information, one obtains
the usual constraint $G_{M\bar M} < 3.0\times 10^{-3} G_F$ for
$(V\pm A)^2$ type interaction~\cite{Willmann:1998gd}. Applying to our case we can put
constraints on the relevant parameters and obtain
\begin{eqnarray}
&&| \frac{A_{d_\U} }{16\sin(d_\U \pi)m^2_M}
\left(\frac{m^2_M}{\Lambda_\U^2}\right)^{d_\U-1}
(c^{\mu e}_{V_{L,R}})^2|
\leq \frac{4.0 \times 10^{-3} G_F}{\sqrt{2}}\;,\nonumber\\
&&| \frac{A_{d_\U} }{16\sin(d_\U \pi)m^2_M}
\left(\frac{m^2_M}{\Lambda_\U^2}\right)^{d_\U}
(c^{\mu e}_{S_{L,R}})^2| \leq \frac{1.2 \times 10^{-2} G_F}{\sqrt{2}}
\;.\label{muonconstr}
\end{eqnarray}
where $G_F$ is the Fermi constant.

\begin{figure}[thb!]
\includegraphics[width=5.5 in]{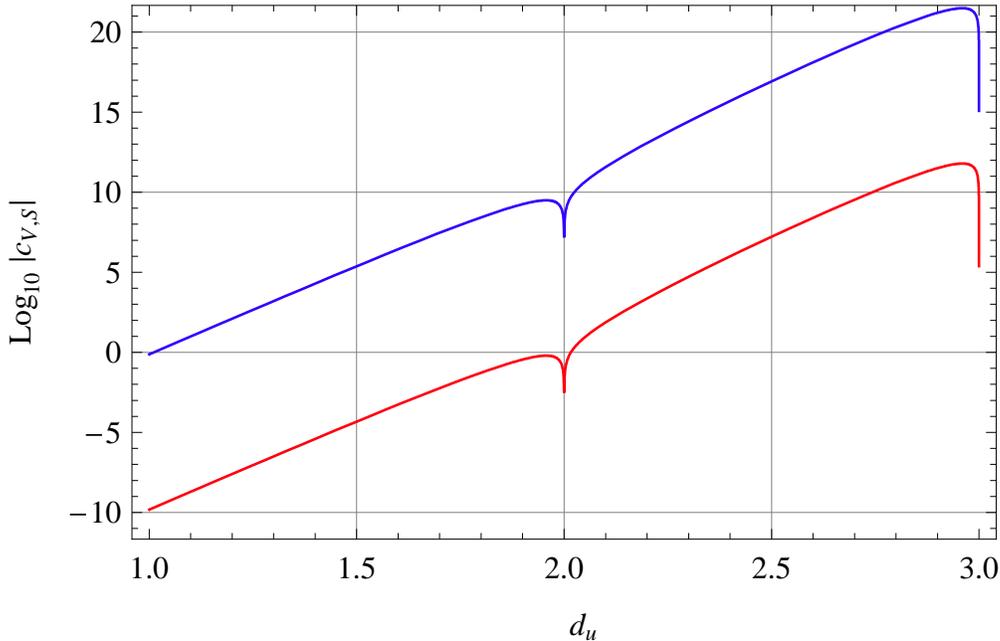}
\caption{\label{muonium} \small Constraints on the coupling
$c_{V,S}$ from muonium-antimuonium oscillation with $\Lambda_\U =
1$ TeV. The upper and lower curves are for $c_S$ and $c_V$
respectively.}
\end{figure}

Using eq.(\ref{muonconstr}), one can obtain constraints on $c_{V,S}$
for given $\Lambda_\U$ and $d_\U$. The constraints for $c_{V,S}$
are shown in Fig.\ref{muonium} for $\Lambda_\U = 1$ TeV. At $d_\U$
equal to integers larger than 1 the contribution to
$\delta$ blows off due to the appearance of $\sin(\pi d_\U)$ in
the denominator of eq.(\ref{matrix}) and therefore one should take
values away from $d_\U$ close to integers. We see that stringent
constraint can be obtained on $c_{V}$ for small $d_\U$.  The
constraint for $c_S$ is weak because the suppression factor of
$m^2_\mu/\Lambda_\U^2$ compared with that for $c_V$.  In general
the constraints are weaker compared with those obtained from meson
and antimeson oscillations since the suppression factors are now
$(m^2_\mu/\Lambda^2_\U)^{d_\U -1}$ and
$(m^2_\mu/\Lambda^2_\U)^{d_\U}$ for vector and scalar unparticle
contributions which are more severe than that for meson-antimeson
oscillation cases.

\section{Discussions and Conclusions}

In our previous discussions, we have assumed that there is no CP
violation in the interactions between unparticles and particles.
We now briefly comment on some implications for CP violation. If
the parameters $\lambda$, $\tilde \lambda$ and $\lambda_Y$ are
complex, CP is violated. There may be chance to have large enough
CP asymmetry for $A_{SL} = (1-|q/p|^4)/(1+|q/p|^4)$ which may be
observed by measuring meson and antimeson semi-leptonic decays.
Let us take $B_d - \bar B_d$ mixing for discussions since there
are a large number of $B_d$ mesons produced at B-factories at KEK
and SLAC and more detailed study could be carried out in the near
future.

In the SM, $A_{SL}$ for $B_d - \bar B_d$ system is predicted to be
very small ($< 10^{-3}$). The reasons for this are two folds:
small $\Gamma_{12}$ and small relative CP violating phase between
$M_{12}$ ad $\Gamma_{12}$. With unparticle interactions,
$\Gamma_{12}$ can be much larger than the SM prediction as shown
before, and with non-zero CP violating phases for $\lambda$,
$\tilde \lambda$ and $\lambda_Y$ relative CP violating phase
between $\Delta m_{12}$ and $\Gamma_{12}$ can be generated. It is
possible to have a sizeable $A_{SL}$. To a good approximation, we
have, $A_{SL} \approx |\Gamma_{12}^\U|\sin\phi /
|M_{12}^{total}|$. Here $\phi$ is a relative CP violating phase
between total $M^{total}_{12}$ and $\Gamma^\U_{12}$ which is
unknown.

Experimentally, $A_{SL}$ is constrained to be~\cite{Yao:2006px}
$-0.0049\pm 0.0038$. Allowing the unparticle contribution to
saturate experimental upper bound on $y_{B_d}$, $A_{SL}$ can
easily reach present constraint. Similar situation
occurs for CP violation in $B_s - \bar B_s$ system. Measurements of
$A_{SL}$ can also provide information about unparticle
interactions.

To summarize, we have studied unparticle effects on particle and
antiparticle oscillations in meson-antimeson, and muonium-antimuonium
systems. We found that unlike usual tree level
contributions to meson oscillations from heavy particle exchange
with small $\Gamma_{12}$, the unparticle may have sizeable
contributions to both $M_{12}$ and $\Gamma_{12}$ due to the fractional
dimension $d_\U$ of the unparticle. Numerically we found that very
stringent constraints on the unparticle and particle interactions
can be obtained. If unparticle effect dominates the contributions
(which may happen in $D^0-\bar D^0$ mixing) to meson mixing
parameters $x$ and $y$, $x/y =\cot(\pi d_\U)$. New constraints on
unparticle and particle interactions can also be obtained using
muonium and antimuonium oscillation data. Unparticle interactions
can also induce large CP violation in meson oscillations.

\vskip 1.0cm \noindent {\bf Acknowledgments}$\,$ The work of
authors was supported in part by the NSC, NCTS and NNSF.


\begin{references}
\bibitem{Georgi:2007ek}
  H.~Georgi,
  Phys.\ Rev.\ Lett.\  {\bf 98}, 221601 (2007)
  [arXiv:hep-ph/0703260];
  H.~Georgi,
  Phys.\ Lett.\  B {\bf 650}, 275 (2007)
  [arXiv:0704.2457 [hep-ph]].

\bibitem{collider}
  K.~Cheung, W.~Y.~Keung and T.~C.~Yuan,
  Phys.\ Rev.\ Lett.\  {\bf 99}, 051803 (2007)
  [arXiv:0704.2588 [hep-ph]];
  P.~J.~Fox, A.~Rajaraman and Y.~Shirman,
  arXiv:0705.3092 [hep-ph];
  N.~Greiner,
  Phys.\ Lett.\  B {\bf 653}, 75 (2007)
  [arXiv:0705.3518 [hep-ph]];
  S.~L.~Chen and X.~G.~He,
  arXiv:0705.3946 [hep-ph];
  P.~Mathews and V.~Ravindran,
  arXiv:0705.4599 [hep-ph].
  M.~Bander, J.~L.~Feng, A.~Rajaraman and Y.~Shirman,
  arXiv:0706.2677 [hep-ph];
  T.~G.~Rizzo,
  arXiv:0706.3025 [hep-ph];
  K.~Cheung, W.~Y.~Keung and T.~C.~Yuan,
  Phys.\ Rev.\  D {\bf 76}, 055003 (2007)
  [arXiv:0706.3155 [hep-ph]];
  T.~Kikuchi and N.~Okada,
  arXiv:0707.0893 [hep-ph];
  D.~Choudhury and D.~K.~Ghosh,
  arXiv:0707.2074 [hep-ph];
  H.~Zhang, C.~S.~Li and Z.~Li,
  arXiv:0707.2132 [hep-ph];
  N.~G.~Deshpande, X.~G.~He and J.~Jiang,
  arXiv:0707.2959 [hep-ph];
  A.~Delgado, J.~R.~Espinosa and M.~Quiros,
  arXiv:0707.4309 [hep-ph].
  M.~x.~Luo, W.~Wu and G.~h.~Zhu,
  arXiv:0708.0671 [hep-ph];
  A.~T.~Alan and N.~K.~Pak,
  arXiv:0708.3802 [hep-ph];
  T.~i.~Hur, P.~Ko and X.~H.~Wu,
  arXiv:0709.0629 [hep-ph];
  S.~Majhi,
  arXiv:0709.1960 [hep-ph];
  M.~C.~Kumar, P.~Mathews, V.~Ravindran and A.~Tripathi,
  arXiv:0709.2478 [hep-ph];
  K.~m.~Cheung, W.~Y.~Keung and T.~C.~Yuan,
  arXiv:0710.2230 [hep-ph].



\bibitem{lowenergy}
  G.~J.~Ding and M.~L.~Yan,
  arXiv:0705.0794 [hep-ph];
  Y.~Liao,
  Phys.\ Rev.\  D {\bf 76}, 056006 (2007)
  [arXiv:0705.0837 [hep-ph]];
  S.~Zhou,
  arXiv:0706.0302 [hep-ph];
  G.~J.~Ding and M.~L.~Yan,
  arXiv:0706.0325 [hep-ph];
  S.~L.~Chen, X.~G.~He and H.~C.~Tsai,
  arXiv:0707.0187 [hep-ph];
  R.~Zwicky,
  arXiv:0707.0677 [hep-ph];
  X.~Q.~Li, Y.~Liu and Z.~T.~Wei,
  arXiv:0707.2285 [hep-ph];
  G.~Bhattacharyya, D.~Choudhury and D.~K.~Ghosh,
  arXiv:0708.2835 [hep-ph].
  A.~B.~Balantekin and K.~O.~Ozansoy,
  arXiv:0710.0028 [hep-ph];
  E.~O.~Iltan,
  arXiv:0710.2677 [hep-ph].


\bibitem{b-mixing}
  M.~Luo and G.~Zhu,
  arXiv:0704.3532 [hep-ph];
   C.~H.~Chen and C.~Q.~Geng,
  arXiv:0705.0689 [hep-ph];
  X.~Q.~Li and Z.~T.~Wei,
  Phys.\ Lett.\  B {\bf 651}, 380 (2007)
  [arXiv:0705.1821 [hep-ph]].
  R.~Mohanta and A.~K.~Giri,
  arXiv:0707.1234 [hep-ph];
  A.~Lenz,
  Phys.\ Rev.\  D {\bf 76}, 065006 (2007)
  [arXiv:0707.1535 [hep-ph]].





\bibitem{flavor1}
  T.~M.~Aliev, A.~S.~Cornell and N.~Gaur,
  arXiv:0705.1326 [hep-ph];
  C.~D.~Lu, W.~Wang and Y.~M.~Wang,
  Phys.\ Rev.\  D {\bf 76}, 077701 (2007)
  [arXiv:0705.2909 [hep-ph]];
  D.~Choudhury, D.~K.~Ghosh and Mamta,
  arXiv:0705.3637 [hep-ph];
  T.~M.~Aliev, A.~S.~Cornell and N.~Gaur,
  JHEP {\bf 0707}, 072 (2007)
  [arXiv:0705.4542 [hep-ph]];
  C.~H.~Chen and C.~Q.~Geng,
  Phys.\ Rev.\  D {\bf 76}, 036007 (2007)
  [arXiv:0706.0850 [hep-ph]];
  C.~S.~Huang and X.~H.~Wu,
  arXiv:0707.1268 [hep-ph];
  R.~Mohanta and A.~K.~Giri,
  Phys.\ Rev.\  D {\bf 76}, 057701 (2007)
  [arXiv:0707.3308 [hep-ph]];
  C.~H.~Chen and C.~Q.~Geng,
  arXiv:0709.0235 [hep-ph];
  G.~j.~Ding and M.~L.~Yan,
  arXiv:0709.3435 [hep-ph];
  T.~M.~Aliev and M.~Savci,
  arXiv:0710.1505 [hep-ph].

\bibitem{longrange}
  Y.~Liao and J.~Y.~Liu,
  arXiv:0706.1284 [hep-ph];
  H.~Goldberg and P.~Nath,
  arXiv:0706.3898 [hep-ph];
  N.~G.~Deshpande, S.~D.~H.~Hsu and J.~Jiang,
  arXiv:0708.2735 [hep-ph];
  S.~Das, S.~Mohanty and K.~Rao,
  arXiv:0709.2583 [hep-ph].




\bibitem{astro}
  H.~Davoudiasl,
  arXiv:0705.3636 [hep-ph];
  S.~Hannestad, G.~Raffelt and Y.~Y.~Y.~Wong,
  arXiv:0708.1404 [hep-ph];
  P.~K.~Das,
  arXiv:0708.2812 [hep-ph];
  D.~Majumdar,
  arXiv:0708.3485 [hep-ph];
  A.~Freitas and D.~Wyler,
  arXiv:0708.4339 [hep-ph];
  L.~Anchordoqui and H.~Goldberg,
  arXiv:0709.0678 [hep-ph];
  J.~McDonald,
  arXiv:0709.2350 [hep-ph].


\bibitem{theory}
  M.~A.~Stephanov,
  Phys.\ Rev.\  D {\bf 76}, 035008 (2007)
  [arXiv:0705.3049 [hep-ph]].
  Y.~Nakayama,
  arXiv:0707.2451 [hep-ph];
  T.~A.~Ryttov and F.~Sannino,
  arXiv:0707.3166 [hep-th];
  M.~Neubert,
  arXiv:0708.0036 [hep-ph];
  Y.~Liao,
  arXiv:0708.3327 [hep-ph];
  I.~Gogoladze, N.~Okada and Q.~Shafi,
  arXiv:0708.4405 [hep-ph].




\bibitem{Okamoto:2005zg}
  M.~Okamoto,
  PoS {\bf LAT2005}, 013 (2006)
  [arXiv:hep-lat/0510113].





\bibitem{Staric:2007dt}
  M.~Staric {\it et al.}  [Belle Collaboration],
  Phys.\ Rev.\ Lett.\  {\bf 98}, 211803 (2007)
  [arXiv:hep-ex/0703036]; K.~Abe {\it et al.}  [BELLE Collaboration],
  arXiv:0704.1000 [hep-ex].
  B.~Aubert {\it et al.}  [BABAR Collaboration],
  Phys.\ Rev.\ Lett.\  {\bf 98}, 211802 (2007)
  [arXiv:hep-ex/0703020].

\bibitem{hfag_charm}Heavy Flavour Average Group (HFAG), http://www.slac.stanford.edu/xorg/hfag/charm/index.html

\bibitem{Yao:2006px}
  W.~M.~Yao {\it et al.}  [Particle Data Group],
  J.\ Phys.\ G {\bf 33}, 1 (2006);

\bibitem{HFAG}
Heavy Flavour Average Group (HFAG) for the 2007 web update of the
Particle Data Group review,
http://www.slac.stanford.edu/xorg/hfag/osc/PDG\_2007/\#DG

\bibitem{LN2006}
  A.~Lenz and U.~Nierste,
  JHEP {\bf 06} (2007) 072
  [arXiv:hep-ph/0612167].

\bibitem{deltamsexp}
  A.~Abulencia {\it et al.}  [CDF Collaboration],
  arXiv:hep-ex/0609040;
A.~Abulencia  [CDF - Run II Collaboration],
 Phys.\ Rev.\ Lett.\  {\bf 97} (2006) 062003
  [arXiv:hep-ex/0606027];
 V.~M.~Abazov {\it et al.}  [D0 Collaboration],
  Phys.\ Rev.\ Lett.\  {\bf 97} (2006) 021802
  [arXiv:hep-ex/0603029].



\bibitem{Willmann:1998gd}
  L.~Willmann {\it et al.},
  Phys.\ Rev.\ Lett.\  {\bf 82}, 49 (1999)
  arXiv:9807011[hep-ex].

\bibitem{formula}
G.~Feinberg and S.~Weinberg,  Phys.\ Rev.\ Lett.\  {\bf 6}, 381 (1961). Phys.\ Rev.\  {\bf 123}, 1439 (1961).
  M.~L.~Swartz,
  Phys.\ Rev.\  D {\bf 40}, 1521 (1989).

\bibitem{Hou:1995dg}
  W.~S.~Hou and G.~G.~Wong,
  Phys.\ Rev.\  D {\bf 53}, 1537 (1996)
  [arXiv:hep-ph/9504311].
  V.~Pleitez,
  Phys.\ Rev.\  D {\bf 61}, 057903 (2000)
  [arXiv:hep-ph/9905406].



\bibitem{Horikawa:1995ae}
  K.~Horikawa and K.~Sasaki,
  Phys.\ Rev.\  D {\bf 53}, 560 (1996)
  [arXiv:hep-ph/9504218].


\end{references}
\end{document}